\documentstyle[prl,aps]{revtex}
\begin{document}
\draft
%\twocolumn
\title{Jaynes principle {\it versus} entanglement}

\author{Ryszard Horodecki\cite{poczta3}, Micha\l{} Horodecki \cite{poczta1}}

\address{Institute of Theoretical Physics and Astrophysics\\
University of Gda\'nsk, 80--952 Gda\'nsk, Poland}

\author{Pawe\l{} Horodecki \cite{poczta2}}

\address{Faculty of Applied Physics and Mathematics\\
Technical University of Gda\'nsk, 80--952 Gda\'nsk, Poland}

\maketitle

\begin{abstract}
We show, by explicit examples, that the Jaynes inference scheme based
on maximization of entropy can produce {\it inseparable} 
states even if there exists a {\it separable} 
state compatible with the measured data. It can lead to problems with 
processing of entanglement. The difficulty vanishes when
one uses inference scheme based on minimization of entanglement.
\end{abstract}

It is well known that quantum mechanics allows to reconstruct completely a
state of the quantum mechanical system from mean values of complete
system of observables measured on the ensemble of identically
prepared systems \cite{reconstr}. By complete set of observables
\cite{reconstr} one means the maximal set of
linearly independent observables where the trivial observable represented
by identity operator is excluded.
In practice we often deal with situations when the state of the system is
unknown and only mean values $a_i$, $(i=1,\ldots,p)$ of some {\it incomplete}
set of observables $\{A_i\}_{i=1}^p$ are available from experiments i.e.
\begin{equation}
{\rm Tr}\varrho A_i\equiv \langle A_i\rangle=
a_i \quad i=1,\ldots,p.
\label{wiezy}
\end{equation}
Then of course there can be many states which are in agreement with
the measured data. It involves the problem of estimation of the state on the
basis of the exact mean values of given observables. According to the
maximum entropy principle \cite{Gibbs,Jaynes,Neumann} we have to
choose from a set of states $\varrho$ which fulfil the constraint
(\ref{wiezy}) the most probable (or representative) state $\varrho_J$ which
maximizes the von Neumann entropy
\begin{equation}
S(\varrho)=-{\rm Tr} \varrho\ln\varrho.
\label{entropy}
\end{equation}
Then the representative state $\varrho_J$ is given by \cite{Jaynes}
\begin{equation}
\varrho_J={Z(\bbox{\lambda)}^{-1}} e^{-\sum_{i=1}^p \lambda_i A_i},
\label{Jaynes}
\end{equation}
where $Z(\bbox{\lambda})={\rm Tr} e^{-\sum_{i=1}^p \lambda_k A_k}$ is the
partition function and the vector $\bbox{\lambda}(a_1,\ldots,a_p)$ is uniquely
determined by the vector $\bbox a=(a_1,\ldots,a_p)$
\begin{equation}
-{\partial\ln  Z(\bbox{\lambda})\over \partial \lambda_i}=a_i,\quad
i=1,\ldots,p.
\end{equation}

The above maximum-entropy principle (or Jaynes principle) was applied
for
partial reconstruction of pure and mixed states of many different systems
\cite{Grandy}. In particular it allowed to interpret quantum statistical
mechanics as a special type of statistical inference \cite{Jaynes}
based on the entropic criterion.

The Jaynes principle is the most rational
inference scheme in the sense that it does not permit to draw
any conclusions unwarranted by the experimental data.  However, this
argument making the principle plausible does not actually prove it
\cite{Wehrl}.
In this context one can ask the question: is the entropic criterion
universal?
Surprisingly, as we will show in this paper there {\it are} situations
where the Jaynes principle fails. This concerns compound quantum systems,
which have recently attracted much attention due to the new
phenomena such as quantum teleportation \cite{Bennett_tel},
quantum dense coding \cite{dense} or quantum cryptography \cite{Wiesner}.

In all the above effects the most
important characteristics of state is entanglement (or inseparability)
\cite{insep}.
%$$$$$$$$$$
Suppose we need the entanglement to deal with one of these effects,
having, however, the compound system in unknown state and some
{\it incomplete} data of type (\ref{wiezy}). Then, usually,
to proceed further, we must somehow estimate the state of the system
from the data. But what scheme of inferring can be used in
this case?
The fact that we need the entanglement for our purposes
imposes a basic condition on possible inference schemes.
Namely they certainly should not give us inseparable
estimated state if only theoretically there exists a separable state compatible
with the measured data. Otherwise it may happen that we
get into troubles trying to use the entanglement we inferred to be present,
when in fact there is no entanglement at all!
Further it will be shown that the entropic criterion does not
protect us from such situations.

Now, a basic question arises: how to check whether
the given constraints could (could not) be satisfied by a separable state?
A natural way is performing
{\it minimization of entanglement}. Clearly the latter must be somehow
quantified. To this end one uses the so-called measures of entanglement  which
vanish for separable states (the latter represent no entanglement) \cite{miary}.
Hence a reasonable inference scheme should involve minimization of a chosen
measure of entanglement. Then one can be sure that if the data could be
produced by a separable state then the estimated state would also be separable.

In this paper we propose the inference scheme based on the entanglement
criterion. The crux of the scheme is minimization of entanglement.
%Then one can be sure
%that if the data could be produced by a separable state then the estimated
%state would also be separable.
This procedure is followed by maximization of entropy which
ensures uniqueness of the resulting state.
%In order to ensure uniqueness of the resulting
%state, the minimization of entanglement is followed by maximization of
%entropy.
We  present explicit
examples of data for which
the minimum entanglement state is {\it separable}, while the state obtained
by means of Jaynes principle is {\it inseparable}.
This occurs even if the
mean values come from measurements made by distant observers who can only
exchange classical information.

Let us make some remarks concerning the procedure of minimization of
entanglement. Note first that since the  entanglement measures
vanish for separable states
%(the latter ones represent no entanglement)
then
they  cannot be strictly convex state functions.
As a consequence, under a given
set of constraints of type (\ref{wiezy}) the state of minimum entanglement does
not need to be  unique. To overcome the difficulty, we propose to
maximize the von Numann entropy {\it after} minimization of entanglement.
Such a procedure produces {\it unique representative state}
\cite{unique}. We shall denote it by $\varrho_E$ where $E$ is the used measure
of entanglement.

In our analysis we will use two measures: entanglement of formation
$E_f$ \cite{huge} and relative entropy entanglement $E_r$ \cite{Knight}.
Both of them are calculated for the two spin-$1\over2$ states diagonal
in the Bell basis \cite{Ef}
given by \cite{Mann}
\begin{eqnarray}
&\psi_{1\atop(2)}\equiv\Phi^{\mp}={1\over\sqrt2}(|\uparrow\uparrow\rangle\mp|\downarrow\downarrow
\rangle)\nonumber\\
&\psi_{3\atop(0)}\equiv\Psi^{\pm}={1\over\sqrt2}(|\uparrow\downarrow\rangle\pm|\downarrow\uparrow
\rangle).
\label{baza}
\end{eqnarray}
In this case both the measures   depend only   on the largest eigenvalue $F$
of a given state and
are increasing functions  of $F$ \cite{huge,Knight}
\begin{equation}
E_f=H({1\over2}+\sqrt{F(1-F)}), \quad \quad E_r=\ln2-H(F)
\label{measures}
\end{equation}
for $F>{1\over2}$ and $E_r=E_f=0$ otherwise;
here $H(x)=-x\ln x-(1-x)\ln(1-x)$.
Therefore if the state of minimum entanglement
is diagonal in the Bell basis it is of the {\it same} form for both measures (i.e.
$\varrho_{E_f}=\varrho_{E_r}$).

Here we shall illustrate a fundamental difference between the entropic
criterion and entanglement criterion of inference by means
of examples involving
constraints  which are invariant under measurement in the Bell basis. They are
characterized by the following condition: any state which fulfils
the constraints would also satisfy them if subjected to measurement in the Bell
basis. Such constraints will be further called {\it Bell constraints}.
It turns out that for the Bell constraints the number of the state parameters
which are to be varied within the minimization procedure can be considerably reduced.
This follows from the
following lemma.

{\bf Lemma.-} {\it For  the Bell constraints the
representative state $\varrho_E$  (where $E=E_f$ or $E=E_r$) is diagonal
in the Bell basis.}

{\bf Proof.-}
To prove the lemma  we note two important properties of the measurement in
the Bell basis (\ref{baza}). Namely, such a measurement
(i) does not increase entanglement (for the considered measures),
(ii) does not decrease entropy.
In other words  for any state $\varrho$ we have
\begin{equation}
E(\varrho_B)\leq E(\varrho),\quad S(\varrho_B)\geq S(\varrho)
\end{equation}
where $\varrho_B$ is the state resulting from $\varrho$ after performing
measurement in Bell basis
\begin{equation}
\varrho_B=\sum_{i=0}^3 P^B_i\varrho P^B_i
\end{equation}
with $P_i^B=|\psi_i\rangle\langle \psi_i|$. %being projectors corresponding to the vectors (\ref{baza}).
The first property we prove in Appendix. The second one follows from the fact
that the entropy does not decrease under von Neumann measurement
\cite{Wehrl} (one says that measurement {\it enhances mixing}).

Let us take the state $\varrho_{E}$ which is  representative
under some Bell constraints.
Consider a new state $\varrho_E^B$ given by
\begin{equation}
\varrho_E^B=\sum_i P_i^B\varrho_E P_i^B.
\end{equation}
By definition of the Bell constraints $\varrho_E^B$ also satisfies
them.
According to the properties (i) and (ii) we have $E(\varrho_E^B)\leq E(\varrho_E)$
and $S(\varrho_E^B)\geq S(\varrho_E)$. But as $\varrho_E$ is the
representative state, then no other state satisfying the constraints
can be less entangled, hence $E(\varrho_E^B)=E(\varrho_E)$.
As the state $\varrho_E$ is unique then  among the
states of minimum entanglement no other state can have entropy
greater than or equal to $\varrho_E$. %Hence, the state $ \varrho_E^B$ is
%identical with $\varrho_E$  i.e. $\varrho_E^B=\varrho_E$.
In result we have $\varrho_E^B=\varrho_E$.
But this means that the state $\varrho_E$ does not change under the
measurement in Bell basis. This is possible if and only if
$\varrho_E$ is {\it diagonal} in this basis. This ends the proof of the
lemma.

From the lemma it follows that  for the Bell constraints one can perform
the procedure
of minimization of entanglement (and subsequent maximization of entropy)
{\it only} over the Bell diagonal states and in this way would  obtain the
same result as
if the procedure were performed over the {\it whole} set of states satisfying
the constraints.  Then, by formulas (\ref{measures}) the minimization of
entanglement reduces for both measures to the minimization of the largest
eigenvalue of the states, producing then the same representative state
$\varrho_E$. In particular the formulas imply that the latter is
inseparable if and only if the eigenvalue is greater than $1\over2$.

Now we are in position to illustrate the difference between the two
inference schemes.
First  consider the Bell-CHSH observable \cite{Bell}
$B=\sqrt{2}(\sigma_x\otimes\sigma_x+\sigma_z \otimes \sigma_z)=
2\sqrt{2}(|\Phi^+\rangle\langle\Phi^+|-|\Psi^-\rangle\langle\Psi^-|)$
with the mean value
\begin{equation}
{\rm Tr}\varrho B=b,\quad 0\leq b\leq 2\sqrt2
\label{wiez}
\end{equation}
(i.e. we have only one constraint).
Let us first find the representative state $\varrho_E$.
Of course, as $B$ is diagonal in
Bell basis, then it forms Bell constraints. Indeed for any state $\varrho$
we have
\begin{equation}
b={\rm Tr} \varrho B ={\rm Tr} \bigl(\varrho\sum_i P^B_i B P^B_i\bigr)=
{\rm Tr} \bigl(\sum_i P^B_i \varrho P^B_i B\bigr).
\end{equation}
Hence the state  after measurement still satisfies the constraints.
Then we need to minimize
the largest eigenvalue of the state of the form
\begin{equation}
\varrho=
p_1|\Phi^+\rangle\langle\Phi^+|+
p_2|\Psi^-\rangle\langle\Psi^-| +
p_3|\Psi^+\rangle\langle\Psi^+|  +
p_4|\Phi^-\rangle\langle\Phi^-|,
\label{stan}
\end{equation}
where $\sum_i p_i=1$, $p_i\geq0$ and $p_1-p_2={b\over2\sqrt2}$. Note that if
$b\leq\sqrt2$ then for $p_2={1\over2}-{b\over2\sqrt2}$ the state is separable
as then the largest eigenvalue is $p_1={1\over2}$ (we will not calculate the
state $\varrho_E$ in this case).
For $b>\sqrt2$ the state (\ref{stan}) is always inseparable as
$p_1>{1\over2}$. The latter is minimal if $p_2=0$. Then we obtain the family
of states with minimal entanglement of the form
\begin{equation}
\varrho=
{b\over2\sqrt2}|\Phi^+\rangle\langle\Phi^+|+
p_3|\Psi^+\rangle\langle\Psi^+|+
p_4|\Phi^-\rangle\langle\Phi^-|,
\end{equation}
Subsequently, maximizing the von Neumann entropy we obtain
the representative state $\varrho_E$ of the form
\begin{equation}
\varrho_E=
{b\over2\sqrt2}|\Phi^+\rangle\langle\Phi^+|+
({1\over2}-{b\over2\sqrt2})
\left(|\Psi^+\rangle\langle\Psi^+|+
|\Phi^-\rangle\langle\Phi^-|\right), for b>\sqrt{2}.
\end{equation}
Thus we have shown that under the constraint (\ref{wiez}) the representative state
$\varrho_E$ is
separable for $b\leq \sqrt2$ and inseparable for $\sqrt2 < b\leq2\sqrt2$.

Let us now apply the Jaynes inference scheme to the same data.
Then the Jaynes state, calculated directly by use of the formula (\ref{Jaynes})
is given by
\begin{eqnarray}
&\varrho_J=
{1\over4}\bigl[
(1-{b\over\sqrt2}+{b^2\over 8})|\Phi^+\rangle\langle\Phi^+|+\nonumber\\
&(1+{b\over\sqrt2}+{b^2\over 8})|\Psi^-\rangle\langle\Psi^-|+
(1-{b^2\over8})\left(|\Psi^+\rangle\langle\Psi^+|+
|\Phi^-\rangle\langle\Phi^-|\right)
\bigr].
\end{eqnarray}
The above state is inseparable for $b>4-2\sqrt2$. Then in the range
$4-2\sqrt2<b\leq\sqrt2$ the Jaynes inference produces {\it inseparable}
state while the minimum entanglement state is {\it separable}.
Although we used some particular entanglement measures, the result is {\it
general}: the property that the state $\varrho_E$ is
separable (inseparable) does not depend on the type of measure $E$.

One could think that this difference between the two types of inference
is due the fact that the used observable is nonlocal, i.e. it cannot
be measured itself without interchange of quantum information  between
the observers.
If the measurements are performed locally then
the mean value of Bell-CHSH observable is not the only measured quantity as
we simultaneously obtain the mean values of the product observables which add
up to the observable. Moreover,  by measuring the product observable
we gain an additional information. Indeed, if the correlations are
measured, the marginal distributions are also obtained.
Then consider the following data which could
be obtained by distant observers (who can communicate only by means of
classical bits)
\begin{eqnarray}
&\langle\sqrt{2}\sigma_x\otimes \sigma_x\rangle=
\langle\sqrt{2}\sigma_z\otimes \sigma_z\rangle={b\over2},\nonumber\\
&\langle\sigma_x\otimes I\rangle=
\langle\sigma_z\otimes I\rangle=
\langle I\otimes \sigma_x\rangle=
\langle I\otimes \sigma_z\rangle=0.
\end{eqnarray}
One can check that these are again Bell constraints. Then
the state $\varrho_E$ can be derived similarly as in the previous
case. Here we obtain the {\it same} results as in the case of the
Bell-CHSH observable measured alone (except
 that for $b\leq\sqrt2$ the state
$\varrho_E$ may be of different form   still however being separable).

Finally it is interesting to consider the projector $P_0$
corresponding  to the singlet state vector $\Psi^-$, which is
a manifestly nonlocal observable. One can check
that here $\varrho_E=\varrho_J$ for any mean value $F={\rm Tr}\varrho P_0$
and both the states are equal to a suitable Werner state
\cite{Werner,Popescu}. % (mixture of the completely chaotic state with the singlet state).
This involves an  interesting problem:
for which type of constraints the Jaynes scheme fails?
However, it goes beyond the scope of this paper.

It should be mentioned here  that the problem of the entanglement processing
with incomplete data appeared implicitly in the context of the
protocols of entanglement distillation.
Indeed, the first proposed distillation scheme \cite{Bennett_pur}
is based on information about the state given  just by the projector $P_0$.
In result, in contrast with more general schemes which involve full knowledge
about the state \cite{pur}, it works only for $F>{1\over2}$. In the
present context, this appears to be a consequence of the fact that  the
minimum entanglement state for $F\leq {1\over2}$ is separable.
Now if the real state is in fact inseparable, we must gain some more
information (i.e. to increase the number of observables) to be able to
distill the state.

Finally, one can ask what is the place of the two inference
schemes (the entropic one and the entanglement one) in quantum
communication theory. It seems that they are in a way complementary.
As the quantum noisy channels are usually described in terms of entanglement
\cite{huge,SL}, the scheme proposed in this paper could be a suitable tool
for estimation of parameters of quantum channels.
On the other hand, the capacity of a quantum
source is described by von Neumann entropy \cite{Schumacher}. Thus
the Jaynes principle is here the natural scheme.

This work is supported by Polish Committee for Scientific Research,
Contract No. 2 P03B 024 12.

\begin{appendix}
\section{}

We will show here that $E(\varrho_B)\leq E(\varrho)$ where
$\varrho_B=\sum_iP_i^B\varrho P_i^B$, $E=E_f$ or $E_r$.

To prove the above, note first that  $\varrho_B $ is diagonal in the Bell
basis i.e.  it is of the form $\varrho_B=\sum_i\lambda_i P_i^B$. Let $\lambda_k$
be the largest eigenvalue. Consider the state
$\varrho_B'=
I\otimes \sigma_k\varrho_B I\otimes \sigma_k$
(where $\sigma_{1, 2, 3}$ are Pauli matrices, $\sigma_0=I$ )
which is still diagonal in the Bell basis. Clearly we have
$\varrho_B'=\sum_i P_i^B\varrho' P_i^B$, where $\varrho'=
I\otimes \sigma_k\varrho I\otimes \sigma_k$. As we have
$I \otimes \sigma_{k} P_k \varrho I\otimes \sigma_k=P_0$,
the largest eigenvalue of $\varrho_B'$ i.e. $\lambda_k$ corresponds
now to the singlet projector $P_0$. Now, we will use the
``twirling'' operation \cite{Bennett_pur} i.e. random unitary transformation
of the form $U\otimes U$.The twirling converts any state $\tilde\varrho$
into the Werner state \cite{Werner,Bennett_pur} $\varrho_W={\rm
Twirl}(\tilde\varrho)$ given by
\begin{equation}
\varrho_W=FP_0+({1-F\over3})\bigl( P_1+P_2+P_3\bigr),
\end{equation}
where $F={\rm Tr} \tilde{\varrho} P_0$. Now the equality
${\rm Tr} \varrho_B'P_0={\rm Tr}\varrho' P_0$ implies
${\rm Twirl}(\varrho')={\rm Twirl}(\varrho_B')$. Moreover, since for Bell diagonal state
both entanglement measures depend only on the largest eigenvalue, we get
$E({\rm Twirl}(\varrho_B'))=E(\varrho_B')$. Subsequently, as twirling involves
only local quantum operations and classical communication we have
$E({\rm Twirl}(\varrho'))\leq E(\varrho')$.  %$E(\varrho_B')\leq E(\varrho')$.
Finally we obtain
\begin{equation}
E(\varrho_B')=E({\rm Twirl}(\varrho_B'))=
E({\rm Twirl}(\varrho'))\leq E(\varrho')
\end{equation}
As entanglement measures are invariant under product unitary transformation
(in particular under $I \otimes \sigma_k$ ones) we have also
$E(\varrho_B)\leq E(\varrho)$.

\end{appendix}


\begin{references}
\bibitem[*]{poczta3} E-mail address: fizrh@univ.gda.pl
\bibitem[**]{poczta1} E-mail address: michalh@iftia.univ.gda.pl
\bibitem[***]{poczta2} E-mail address: pawel@mif.pg.gda.pl
\bibitem{reconstr}
W. Band and J. L. Park, Found. Phys. {\bf 1}, 133 (1970);
Am. J. Phys. {\bf 47}, 188 (1979);
\bibitem{Gibbs}
J. W. Gibbs, Elementary Principles in Statistical Mechanics (Yale University
Press, 1902)
\bibitem{Jaynes}
E. Jaynes, Phys. Rev. {\bf 108}, 171 (1957); {\it ibid} {\bf 108}, 620 (1957);
Am. J. Phys. {\bf 31}, 66 (1963).
\bibitem{Neumann}
W. M. Elsasser, Phys. Rev. {\bf 52}, 987 (1937);
J. von Neumann, Mathematical Foundations of Quantum Mechanics, Princeton
1955, Princeton Univ. Press; R. S. Ingarden and K.
Urbanik, Bull. Acad. Polon. Sci. {\bf 9}, 313 (1961).
\bibitem{Grandy}
See e.g. V. Bu\v{z}ek, G. Drobn\'y, G. Adam, R. Derka and P. L. Knight,
J. Mod. Opt. {\bf 44}, 2607 (1997);
for an extensive presentation of the use of Jaynes principle see
W. T. Grandy, Am. J. Phys {\bf 65}, 466 (1997).
\bibitem{Wehrl}
A. Wehrl, Rev. Mod. Phys. 50 (1978) 221.
\bibitem{Bennett_tel}
C. Bennett, G. Brassard, C. Crepeau, R. Jozsa, A. Peres and W. K. Wootters,
Phys. Rev. Lett.  {\bf 70}, 1895  (1993).
\bibitem{Wiesner}
S. Wiesner SIGACT News {\bf 15}, 78 (1983);
C. H. Bennett and G. Brassard, 1984, Proceedings of the IEEE International
Conference on Computers  (New York: IEEE), p. 175;
Ekert A. Phys. Rev. Lett. {\bf 67},  661 (1991).
\bibitem{dense}
C. H. Bennett and S. J. Wiesner Phys. Rev. Lett. {\bf 69} 2881 (1992).
\bibitem{insep}
A mixed state is separable (inseparable) if it can (cannot)
be written as a mixture of product states (they can be taken to be pure)
$\varrho_{sep}=\sum_ip_i|\psi_i\otimes
\phi_i\rangle\langle\psi_i\otimes\phi_i|$ \cite{Werner} (see
also P. Horodecki Phys. Lett. A {\bf 232}, 233 (1997)).
\bibitem{Werner}
R. F. Werner, Phys. Rev. A {\bf 40}, 4277 (1989);
\bibitem{miary}
For a review of entanglement measures and their properties see
V. Vedral and M. B. Plenio, Report No. quant-ph/9707035.
In particular, entanglement measure is required to have two
important properties: it must be invariant under product unitary
transformations and it cannot increase under the operation consisting of
local quantum operation and classical communication. The latter
property implies that entanglement measures are convex functions.
\bibitem{unique}
Since
the constraints are linear, and entanglement measures are convex fuctions
then the set of states with minimal entanglement is convex. Consequently,
as entropy is a {\it strictly} convex function then maximizing entropy
over the convex set we obtain  unique state.
\bibitem{huge}
C. H. Bennett, D. P. Di Vincenzo, J. Smolin and
W. K. Wootters, Phys. Rev. A {\bf 54},  3814 (1997).
\bibitem{Knight}
V. Vedral, M. B. Plenio, K. Jacobs and P. L. Knight, Phys. Rev. Lett.
{\bf 78}, 2275 (1997).
\bibitem{Ef}
There is analytical formula for entanglement of formation of
any two spin-$1\over2$ state, see
S. Hill and W. K. Wooters, Phys. Rev. Lett. {\bf 78}, 5022 (1997);
W. K. Wooters, Report No. quant-ph/
\bibitem{Mann}
S.L. Braunstein, A. Mann, M. Revzen, Phys. Rev. Lett.
{\bf 68}, 3259  (1992).
\bibitem{Bell}
We use here a special case \cite{Mann} of the Bell-CHSH observable
\cite{CHSH}
$B_{CHSH}=a\otimes b+a\otimes b' +a'\otimes b -a'\otimes b'$
where $a,b,a',b'$ are dichotomic observables.
\bibitem{CHSH}
J. F. Clauser, M. A. Horne, A. Shimony and R. A. Holt, Phys. Rev. Lett.  {\bf
23}, 880 (1969);
\bibitem{Popescu}
S. Popescu,  Phys. Rev. Lett.  {\bf 72},  779 (1994).
\bibitem{Bennett_pur}
C. H. Bennett, G. Brassard, S. Popescu, B. Schumacher, J. Smolin and
W. K. Wootters, Phys. Rev. Lett. {\bf 76}, 722 (1996).
\bibitem{pur}
M. Horodecki,  P. Horodecki and R. Horodecki, Phys. Rev. Lett.
{\bf 78} (1997) 574.
\bibitem{SL}
B. Schumacher and M. A. Nielsen, Phys. Rev. A {\bf 54}, 2629 (1996);
S. Lloyd Phys. Rev. A {\bf 54}, 1613 (1997).
\bibitem{Schumacher}
B. Schumacher, Phys. Rev. A {\bf 51}, 2738 (1995); see also
R. Jozsa and B. Schumacher, J. Mod. Opt. {\bf 41}, 2343 (1994);
H. Barnum, Ch. Fuchs, R. Jozsa and B. Schumacher,
Phys. Rev. A {\bf 54}, 4707 (1996).
\end{references}
\end{document}